\documentclass[runningheads]{llncs}
\usepackage{graphicx}

\usepackage{tikz}
\usepackage{comment}
\usepackage{amsmath,amssymb} 
\usepackage{color}

\usepackage{graphicx}
\usepackage{amsmath,amssymb} 
\usepackage{times}
\usepackage{epsfig}
\usepackage{graphicx}
\usepackage{amsmath}
\usepackage{amssymb}
\usepackage{colortbl}
\usepackage{relsize}
\usepackage{multirow}
\usepackage{float}

\usepackage[export]{adjustbox}

\usepackage[pagebackref=true,breaklinks=true,colorlinks,bookmarks=false]{hyperref}

\usepackage[accsupp]{axessibility}  

\usepackage[width=122mm,left=12mm,paperwidth=146mm,height=193mm,top=12mm,paperheight=217mm]{geometry}

\begin{document}
\pagestyle{headings}
\mainmatter
\def\ECCVSubNumber{1115}  

\title{Residual Aligner Network}

\titlerunning{Residual Aligner Network}
%
\author{Jian-Qing Zheng\inst{1,2} \and
Ziyang Wang \inst{3} \and
Baoru Huang\inst{4} \and Ngee Han Lim\inst{1} \and Bart{\l}omiej W. Papie{\.z}\inst{2,5}}
\authorrunning{J.-Q. Zheng et al.}
%
\institute{The Kennedy Institute of Rheumatology, University of Oxford, U.K.
\and
Big Data Institute, University of Oxford, U.K.
\and
Department of Computer Science, University of Oxford, U.K.
\and
Department of Surgery and Cancer, Imperial College London
\and
Nuffield Department of Population Health, University of Oxford, UK
\email{\{jianqing.zheng@kennedy,bartlomiej.papiez@bdi\}.ox.ac.uk}}

\maketitle

\begin{abstract}

Image registration is important for medical imaging, the estimation of the spatial transformation between different images.
Many previous studies have used learning-based methods for coarse-to-fine registration to efficiently perform 3D image registration. The coarse-to-fine approach, however, is limited when dealing with the different motions of nearby objects.  Here we propose a novel Motion-Aware (MA) structure that captures the different motions in a region. The MA structure incorporates a novel Residual Aligner (RA) module which predicts the multi-head displacement field used to disentangle the different motions of multiple neighbouring objects. Compared with other deep learning methods, the network based on the MA structure and RA module achieves one of the most accurate unsupervised inter-subject registration on the 9 organs of assorted sizes in abdominal CT scans, with the highest-ranked registration of the veins (Dice Similarity Coefficient / Average surface distance: 62\%/4.9mm for the vena cava and 34\%/7.9mm for the portal and splenic vein), with a half-sized structure and more efficient computation. Applied to the segmentation of lungs in chest CT scans, the new network achieves results which were indistinguishable from the best-ranked networks (94\%/3.0mm). Additionally, the theorem on predicted motion pattern and the design of MA structure are validated by further analysis.
\end{abstract}

\section{Introduction}
\label{sec:intro}
Alignment of two images, also known as image registration  \cite{sotiras2013deformable}, 
is an important task in computer vision applications. 
In medical imaging, image registration enables comparison between different acquisitions over time (longitudinal analysis) or between different types of scanners (multi-modal registration). 


Image alignment can be defined as estimation of the spatial transformation ${\phi}:\mathbb{R}^{n}\to\mathbb{R}^{n}$, represented by a corresponding parameters or a series of displacements denoted by $\phi[\textbf{\textit{x}}]\in \mathbb{R}^{d}$ at the coordinate $\textbf{\textit{x}}\in\mathbb{Z}^{d}$ of a target image $\textbf{\textit{I}}^{\rm t}\in\mathbb{R}^{{n}}$ from a source image  $\textbf{\textit{I}}^{\rm s}\in\mathbb{R}^{n}$, where $n$ is the size of a 3D image defined as $n={H \times W\times T}$, and $d, T, H, W$ denoting the image dimension, thickness, height, and width, respectively.
Originally, image registration was solved as an optimization problem by minimization of a dissimilarity metric $\mathcal{D}$ and a regularization term $\mathcal{S}$:
\begin{equation}
\label{equ:opt_phi}
\hat{\phi}=\underset{\phi}{\mathrm{argmin}}{\big(\mathcal{D}(\phi(\textbf{\textit{I}}^{\rm s}),~\textbf{\textit{I}}^{\rm t})+\lambda\mathcal{S}(\phi,\textbf{\textit{I}}^{\rm t})\big)}
\end{equation}
where $\hat{\phi}$ denotes the estimated spatial transform,
$\lambda$ denotes the weight of the regularization. 
Several methods including Demons \cite{thirion1998image} or Free Form Deformations \cite{rueckert1999nonrigid} have been proposed to solve Eq.~\eqref{equ:opt_phi}, however they can get trapped in the the local optimum and their computational performance is limited due to iterative optimization of highly dimensional, non-convex problem.

More recently, the alignment is performed 
via (convolution) neural networks $\mathcal{R}$ using the feature maps $\textbf{\textit{F}}^{\rm s}$, $\textbf{\textit{F}}^{\rm t}\in\mathbb{R}^{c\times n}$ extracted from $\textbf{\textit{I}}^{\rm s}$ and $\textbf{\textit{I}}^{\rm f}$ respectively, (and $c$ denotes a number of feature channels) by directly regressing (DR) the spatial transformation  \cite{balakrishnan2018unsupervised,mok2020fast}:
\begin{equation}
\phi=\mathcal{R}(\textbf{\textit{F}}^{\rm s},\textbf{\textit{F}}^{\rm t};w)
\end{equation}
with the training process based on minimizing the loss function (e.g. given in Eq.~\eqref{equ:opt_phi}) with the trainable weights $w$ 
($w$ are omitted in the remaining part of the paper to simplify the equation). 
However the direct regression of spatial transformations via convolution neural networks could suffer due to limited capture range of the receptive field of convolution layers when dealing with large or complex motion such as sliding motion.

One solution is to perform coarse-to-fine alignment via multi-scale feature maps or feature pyramid \cite{ranjan2017optical,sun2018pwc,de2019deep,xu2021f3rnet,chang2017clkn,lv2019taking}.
In coarse-to-fine approach, the residual transformation $\varphi_{k}$ between the target feature map $\textbf{\textit{F}}_k^{\rm t}$ and the warped source feature map based on previous level $k-1$ registration $\phi_{k-1}(\textbf{\textit{F}}_k^{\rm s})$ 
is accumulated following the coarse-to-fine image resolution (thus expanding the receptive field from large to small):
\begin{equation}
\label{equ:res_align}
\left\{
\begin{array}{cc}
    \phi_k=\phi_{k-1}\circ\varphi_{k}\\
    \varphi_{k}=\mathcal{R}(\phi_{k-1}(\textbf{\textit{F}}_k^{\rm s}),\textbf{\textit{F}}_k^{\rm t})\\
\end{array}
\right.
\end{equation}
where 
$\circ$ denotes the composition of two spatial transformations, and $\phi_{0}$ is initialized as the identity transform.
However, it has two limitations. 
First the coarse-to-fine strategy in the above-mentioned methods is performed on pyramid representation of feature maps, leading to the limited accessible range of neighbouring pixel's motion, and thus may not estimate accurately different motions of two nearby objects \cite{xiao2006bilateral} or organs \cite{papiez2014implicit}.
Secondly, those spatial transforms from different scales are usually directly combined at each position with equal weight \cite{zhao2019unsupervised,xu2021f3rnet,ranjan2017optical,chang2017clkn}, leading to 
the lack of flexible balancing between similarity measure and deformation rationality. 

Alternatively, Cross Attention (CA) mechanism \cite{vaswani2017attention} is used in \cite{li2021revisiting,sun2021loftr,heinrich2019closing} to obtain the global receptive field and use so-called indicator matrices to quantify the relationship between each pair of pixels from two images, and the usage of multiple indicator matrices is called multi-head. 
However, calculation of the indicator matrix has $\mathcal{O}(n^2)$ computational and memory complexity, which could be prohibitive for 3D image registration.

In this paper, we propose the Residual Aligner Network (RAN) based on a novel Motion-Aware (MA) structure (Fig.~\ref{fig:fcn}) and a new Residual Aligner (RA) module (Fig.~\ref{fig:res_aligner}) for efficient, motion-aware, coarse-to-fine image registration. 
Our contributions are as follows.
\begin{itemize}
    \item A new MA structure employing dilated convolution \cite{chen2018deeplab} with high-resolution feature maps is introduced to benefit the network on predicting different motion pattern (Sec.~\ref{sec:motion-aware});
    \item 
    Our RA module utilizes confidence and multi-head mechanism based on the semantic information of the image (Sec.~\ref{sec:res_aligner});
    \item The above proposed components constitute an novel RAN that performs efficient, coarse-to-fine, motion-aware unsupervised registration achieving state-of-the-art accuracy on publicly available lung and abdomen Computed Tomography (CT) data in Sec.~\ref{sec:experiments};
    \item We also investigate and quantify the capture range (Sec.~\ref{sec:rf}) and motion patterns (Sec.~\ref{sec:sep_motion}) predicted in coarse-to-fine registration by recursively warping feature maps.
\end{itemize}

\subsection{Related Works}
Voxelmorph \cite{balakrishnan2019voxelmorph} is an early deep learning method using a convolution neural network, U-net \cite{ronneberger2015u}, for deformable medical image registration. 
However, the capture range of large motions by the DR based learning methods is usually limited by the receptive field in the convolution networks between two images which thus requires a pre-alignment.

DIRnet \cite{de2019deep} was thus proposed with multi-stage (MS) networks for coarse-to-fine registration with each network trained for the specific resolution and searching range of registration, using B-spline for interpolation on the sparse prediction, but require extra time cost on training. 
Several end-to-end training multi-stage networks \cite{zhao2019unsupervised,hu2018weakly,shen2019networks,zhao2019recursive} were thus proposed for coarse-to-fine image registration by recursively warping images.
However, the sub-network of each stage is fed with the directly warped images but not well processed feature maps, and lack connection between different stages, leading to calculation and parameters consuming on extracting the repeated features. 

To efficiently employed the features, Feature Pyramid (FP) was employed for registration in Dual-PRNet (DPRn) \cite{hu2019dual} for unsupervised registration. Multiple spatial transforms are predicted in multi-scale feature domain, to gradually refine the registration based on a sequence of feature maps extracted from a compacted structure \cite{hu2019dual}.
Furthermore, Edge-Aware Pyramidal Network \cite{cao2021edge} was designed  for unsupervised registration with an extra edge image of the original input to enhance the texture structure features. 
A new bilevel, self-tune framework \cite{liu2021learning} was also proposed for training a pyramidal-based registration network with contextual regularization.
However, they still simply add or concatenate the predictions from multi-scale equally weighted features without quantifying the predictions' reliability via the information density to flexibly balance between the similarity of registered images and rationality of motions.  
\begin{figure}[t!]
\begin{center}
\includegraphics[width=\linewidth]{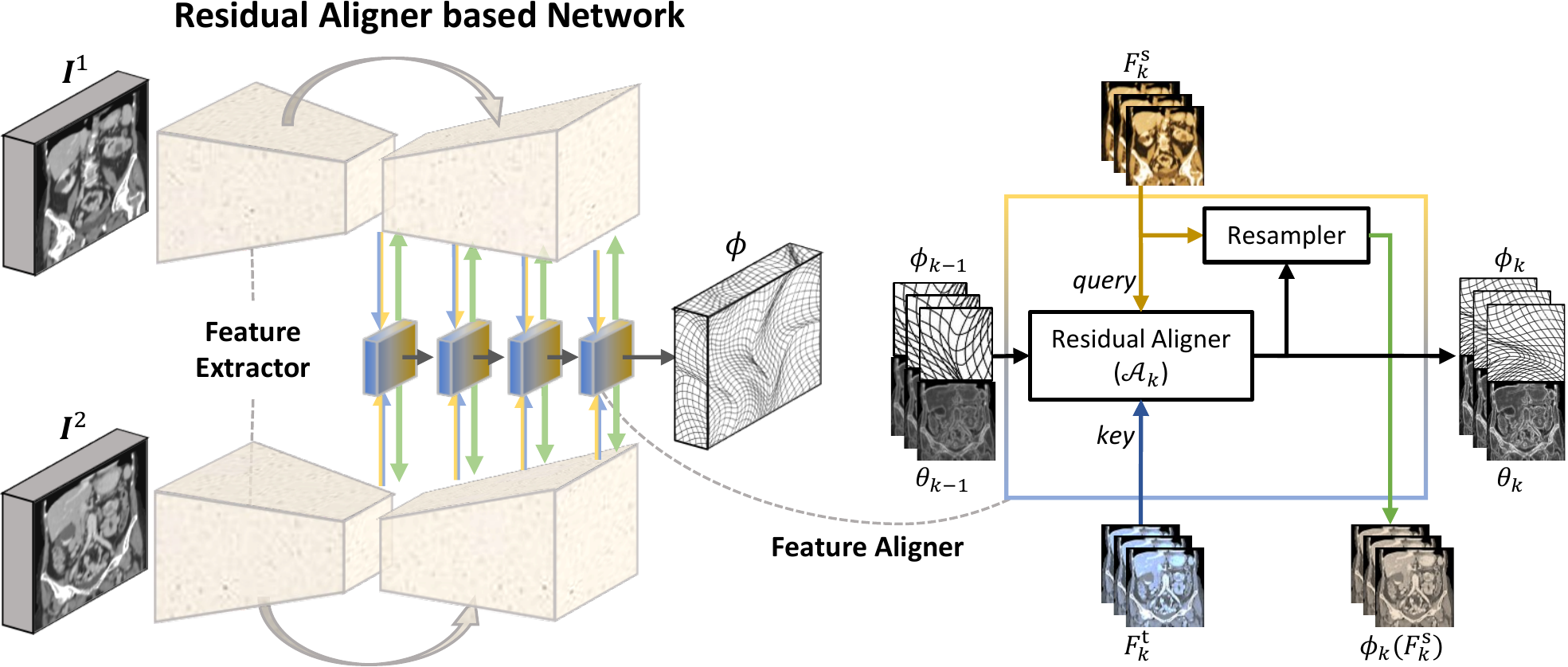}
\end{center}
\vspace{-1.5em}
\caption{The architecture of RAN. Two Motion-Aware feature extractor networks results in feature maps (see more in Fig.~\ref{fig:fcn}) and stacked Residual Aligner modules (see more in Fig.~\ref{fig:res_aligner}) aligns and connects the data streams from the input images.}
\label{fig:ran}
\end{figure}

\section{Network Design for Motion-Aware Structure}
\label{sec:net_design}

\begin{figure}[ht]
\begin{center}
\includegraphics[width=\linewidth]{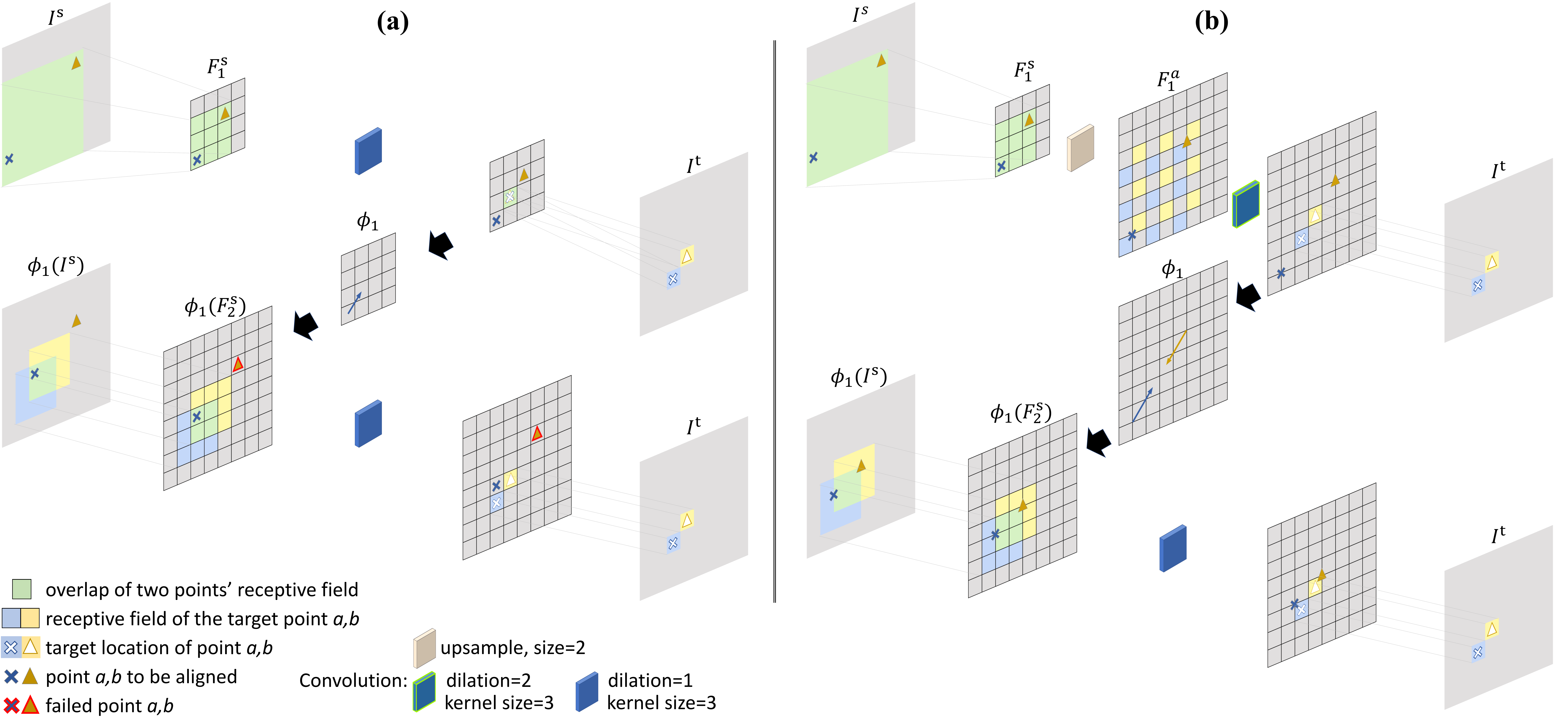}
\end{center}
\vspace{-1.5em}
\caption{Illustration of the problem of coarse-to-fine alignment of two neighbouring points, \textit{a} ($\textcolor{blue}{\times}$) and \textit{b} ($\textcolor{yellow}{\triangle}$), with differing motion. (a) failed capture of point~\textit{b} due to the low resolution feature pyramid. (b) Our proposed solution, utilizing an upsampling layer and dilated convolution in the Motion-Aware structure (Fig.~\ref{fig:fcn}),  while maintaing the same receptive field.}
\label{fig:pyramid_vs_cuboid}
\end{figure}

In this section, we describe the proposed coarse-to-fine image registration network with MA based Feature Extractor to extract the feature maps, and Feature Aligner including stacked RA modules to find the correspondence and estimate the Dense Displacement Field (DDF) as shown in Fig.~\ref{fig:ran}. 
The proposed RA modules are stacked for coarse-to-fine image alignment, and each of them takes two feature maps from two images, align and feed back the warped feature map to enhance the feature extraction, and forward the coarse registration to the next RA module for refining registration, where the details of RA module are described in Sec.~\ref{sec:res_aligner}. 

A pair of Fully Convolution Networks (FCN), with shared weight for efficient training, is used here as the feature extractor to extract two sets of feature maps, $\{\textit{\textbf{F}}_k^{\rm 1}\}_{k=1}^{K}$ and $\{\textit{\textbf{F}}_k^{\rm 2}\}_{k=1}^{K}$, which take turns as the source and target feature maps. The RA module takes the two feature maps from the two streams of FCN, retrieves one (key) on another (query) and then feeds back the aligned feature maps respectively to reinforce the next feature extraction. 
The requirement of RA module for global range attention is described in Sec.~\ref{sec:rf}. The separability of motions is defined in Sec.~\ref{sec:sep_motion} to quantify the bottleneck of separating different motions within a certain range region in the coarse-to-fine registration. A new type of Motion-Aware FCN is proposed to perform alignment at the higher resolution feature maps comparing with Pyramidal FCN in Sec.~\ref{sec:motion-aware}.

\subsection{Capture Range in Coarse-to-fine Registration}
\label{sec:rf}
\begin{definition}
\label{def:motion_diff}
\textbf{(Accessible Motion Range)} :
The radius of capture range of the the $k^{\rm th}$ RA module is defined as the smallest upper bound of its accessible DDF: 
\begin{equation}
\label{equ:cap_rang}
a_k:=\min_{\textbf{\textit{x}}}{\{\sup_{}{({\|\phi_{k}\circ\phi_{k-1}^{-1}[\textbf{\textit{x}}]\|}_{\infty})}\}}
\end{equation}
where $\sup(\cdot)$ denotes the upper bound with varying inputs and the trainable weights of networks, ${\|\cdot\|}_{\infty}$ denotes the L-$\infty$ norm, $\textbf{\textit{x}}$ denotes one coordinate entry of the images or DDFs.
\end{definition}
The accessible motion range can be approximated based on the module's receptive field: $a_k\approx\frac{s_k-1}{2}$, where $s_k$ denotes the original-resolution size of effective receptive field on the input feature maps:
\begin{equation}
\label{equ:rfsize}
{s}_{k}=\underbrace{p_k}_{\rm (i)}\underbrace{(1+2{\|{\textbf{\textit{r}}}_k\|}_1)}_{\rm (ii)}
\quad
\end{equation}
where $\textbf{\textit{r}}_k$ denote the dilation rates of convolution layers in the $k^{\rm th}$ level registration, $p_k$ is the corresponding pool size of the $k^{\rm th}$ feature maps' one pixel on the original image with $p_k\leq{p_{k-1}},\forall{k}>0$, and the convolutions are all assumed with kernel size less or equal than 3 to minimize the computation cost. Thus, the part~(i) and (ii) in Eq.~\eqref{equ:rfsize} are respectively dependent on pool size and dilation.

In the case of global registration on the whole image, the hyper-parameters ${p_1},{\textbf{\textit{r}}_1}$ are set to enable $a_1$ to reach the whole image:
\begin{equation}
\label{equ:global_align}
{p_1}(1+2{\|{\textbf{\textit{r}}}_k\|}_1)\geq2\max(T,H,W)+1, 
\end{equation}
and thus accessible motion range covers the whole image.

\subsection{Motion Separability}
\label{sec:sep_motion}

In the feature pyramid approach, the typical convolution without dilation and the feature pyramid is employed: $r_k=[1~1~\cdots],\forall k\in [0,K]\cap\mathbb{Z}$, $p_k=2^{K-k}$, which fixes the the Eq.~\eqref{equ:rfsize}(ii) and relies on downsampling to enlarge receptive field with only $\mathcal{O}(n)$ complexity to reach the whole image. However, as shown in Fig.~\ref{fig:pyramid_vs_cuboid}(a), the DDF predicted on low-resolution feature map could form a bottleneck of estimated motion's Degree of Freedom (DoF). The only one predicted displacement is occupied by point~\textit{a} instead of point~\textit{b} and thus point~{\textit{b}} is not retrieved until finer resolution but with too smaller receptive field, where points~\textit{a} and \textit{b} could be at the edges of different objects or even just two tiny objects. 
To quantify this phenomenon, we define the separability of the motion prediction:
\begin{definition}
\label{def:motion_diff}
\textbf{(Separability of Predicted Motion)} :
The separability of different motions of the $k^{\rm th}$ RA module is defined as the bottleneck of the upper bound of the difference of its predictable DDF between two locations $\textbf{\textit{x}},\textbf{\textit{y}}\in\mathbb{Z}^{d}$:
\begin{equation}
\label{equ:def_disp_sup}
\Delta_\infty(p):=\min_{\textbf{\textit{x}},\textbf{\textit{y}}}\{\sup_{}({\|\phi[\textbf{\textit{x}}]-\phi[\textbf{\textit{y}}]\|}_{\infty}):{\|\textbf{\textit{x}}-\textbf{\textit{y}}\|}_{\infty}=p\}
\end{equation}
where $p$ denotes the L-$\infty$ distance between the two pixels. 
\end{definition}
The reason of the problem in Fig.~\ref{fig:pyramid_vs_cuboid}(a) is residual registration based on feature pyramid is suffering from the limited range of predicted motion difference with respect to the capture range and the pool size:
\begin{theorem}
\label{thm:reg_consist}
\textbf{(Regional Dependency)} The upper boundary of motion difference is related to $a_k$ and $p_k$:
\begin{equation}
\begin{array}{c}
\forall~\textbf{\textit{x}},\textbf{\textit{y}}\in\mathbb{Z}^{d},~\|\textbf{\textit{x}}-\textbf{\textit{y}}\|_\infty\geq{p_{k''}+2\sum_{k'=k''+1}^{k}{a_{k'}}},\\ \sup_{}(\|\phi_{k}[\textbf{\textit{x}}]-\phi_{k}[\textbf{\textit{y}}]\|_\infty)\geq2\sum_{k'=k''}^{k}{a_{k'}};\\
\exists~\textbf{\textit{x}},\textbf{\textit{y}}\in\mathbb{Z}^{d},~\|\textbf{\textit{x}}-\textbf{\textit{y}}\|_\infty<p_{k''-1}+2\sum_{k'=k''}^{k}{a_{k'}},\\ \sup_{}(\|\phi_{k}[\textbf{\textit{x}}]-\phi_{k}[\textbf{\textit{y}}]\|_\infty)= 2\sum_{k'=k''}^{k}{a_{k'}};
\end{array},
0\leq k''<k
\end{equation}
where $k'',k,\textbf{\textit{x}},\textbf{\textit{y}}$ denote two recursive numbers and two coordinate entries of images/DDFs.
\end{theorem}

Following Theorem~\ref{thm:reg_consist}, $\Delta_{\infty}(p)$ is defined as: 
\begin{equation}
\label{equ:disp_sup}
\Delta_\infty(p)=
\left\{
\begin{array}{llr}
2\sum_{k=1}^{K}{a_{k}},&~p\geq p_{1}+2\sum_{k=2}^{K}{a_{k}}&\\
2\sum_{k'=k}^{K}{a_{k'}},&~p_k+2\sum_{k'=k+1}^{K}{a_{k'}}\leq{p}{<}{p_{k-1}+2\sum_{k'=k}^{K}{a_{k'}}}, &{1<{k}\leq{K}}\\
0,&~{p}<{p_K}&\\
\end{array}
\right.
\end{equation}
to describe the limitation on the multi-objects' motion difference.
More details of Theorem~\ref{thm:reg_consist} are given in Appendix.

\begin{figure}[t]
\begin{center}
\includegraphics[width=\linewidth]{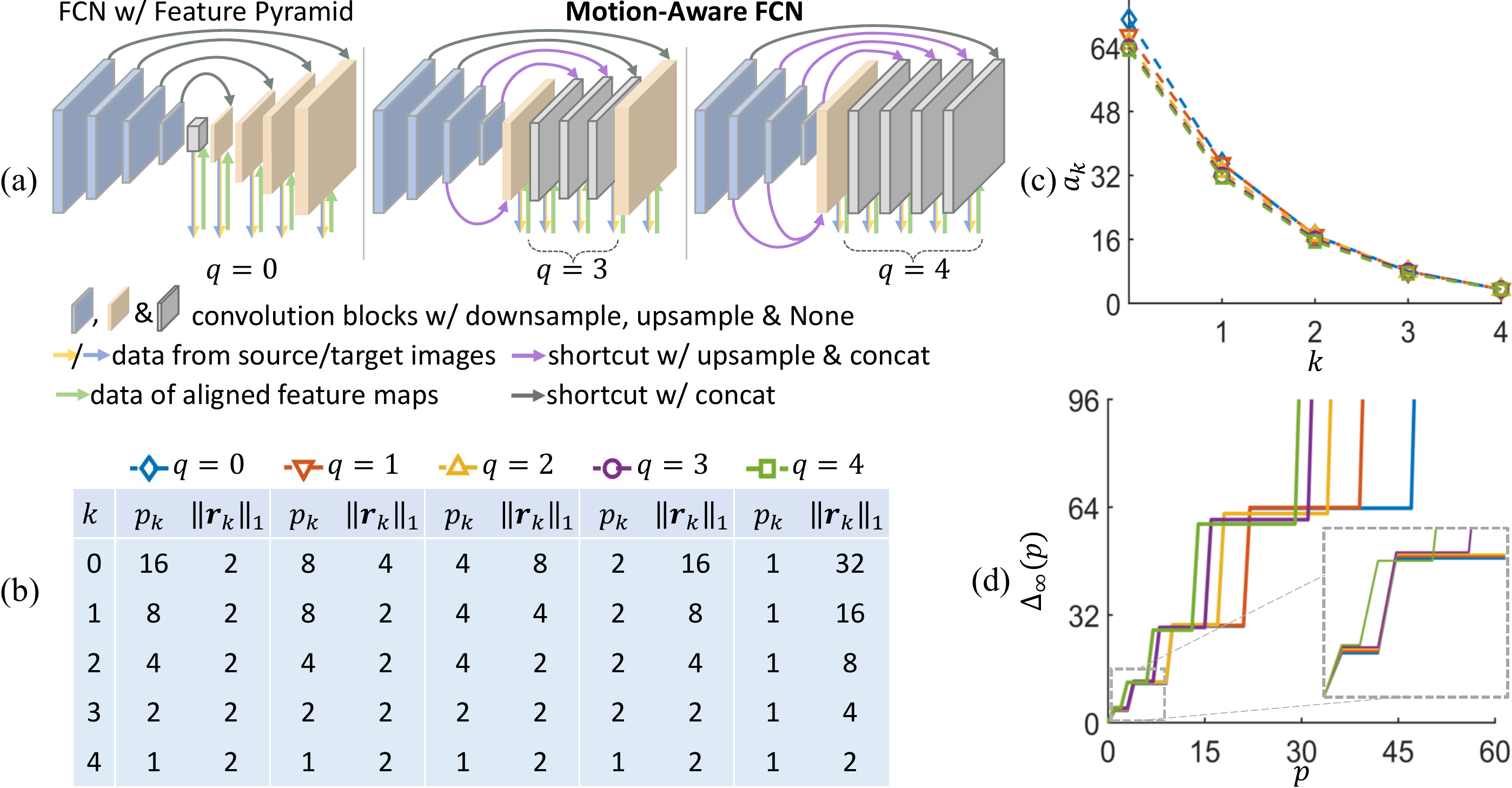}
\end{center}
\vspace{-1.5em}
\caption{The design and theoretical analysis of Fully Convolution Networks (FCN) for feature extraction. (a)  Motion-Aware Structures designed with varying number of motion-aware layers $q$, where the feature maps from encoder part are upsampled and concatenated to the decoder part, (b) with different hyper-parameter setting, showing that a higher $q$, (c) with almost the same $a_k$, achieves higher area under $\Delta_\infty{(p)}$ (d), referring to Eq.~\eqref{equ:cap_rang},\eqref{equ:disp_sup} (unit: pix/vox).}
\label{fig:fcn}
\end{figure}

\subsection{Motion-Aware Structure}
\label{sec:motion-aware}
According to Theorem~\ref{thm:reg_consist}, the smaller pool size releases higher range of motion difference.
Here, we design a new structure, called MA FCN, to achieve high DoF of DDF but still with the same capture range using dilation convolution on upsampled feature maps as shown in Fig.~\ref{fig:fcn}(a). Different from the conventional Feature Pyramid based FCN, the shortcut feature maps from the encoder part are upsampled and concatenated to a specific high-resolution feature map as the input to the decoder part with $p_k=2^{K-q},\forall k\leq{q}$ and $p_k=2^{K-k},\forall q<{k}\leq K$, where $q$ denotes the layer number with MA pattern in the decoder part. The $q$ could be adjusted considering for the balance between the DoF of the predicted DDF and computational cost. The complexity required is $\mathcal{O}(n\log(n))$ using fully MA-layer structure $q=K$ and is still $\mathcal{O}(n)$ using fully feature pyramid $q=0$. To keep the same receptive field of MA structure as FP structure, the dilation rate is set to ${\|{\textbf{\textit{r}}}_k(q>0)\|}_1\geq2^{q-k}{\|{\textbf{\textit{r}}}_k(q=0)\|}_1,\forall k\leq q$ as suggested by Eq.~\eqref{equ:cap_rang} and Eq.~\eqref{equ:rfsize}. As shown in Fig.~\ref{fig:pyramid_vs_cuboid}(b), with the same receptive field, the MA structure releases the higher resolution before alignment and thus avoid loss of the DoF of DDF.
The capture ranges and the difference ranges of DDF for varying setting are illustrated in Fig.~\ref{fig:fcn}(b)(c)(d) based on the calculation of Eq.~\eqref{equ:cap_rang} and \eqref{equ:disp_sup}, where the new design achieves larger area under $\Delta_\infty{(p)}$ with almost the same $a_k$.

\begin{figure*}[t]
\begin{center}
\includegraphics[width=1.\linewidth]{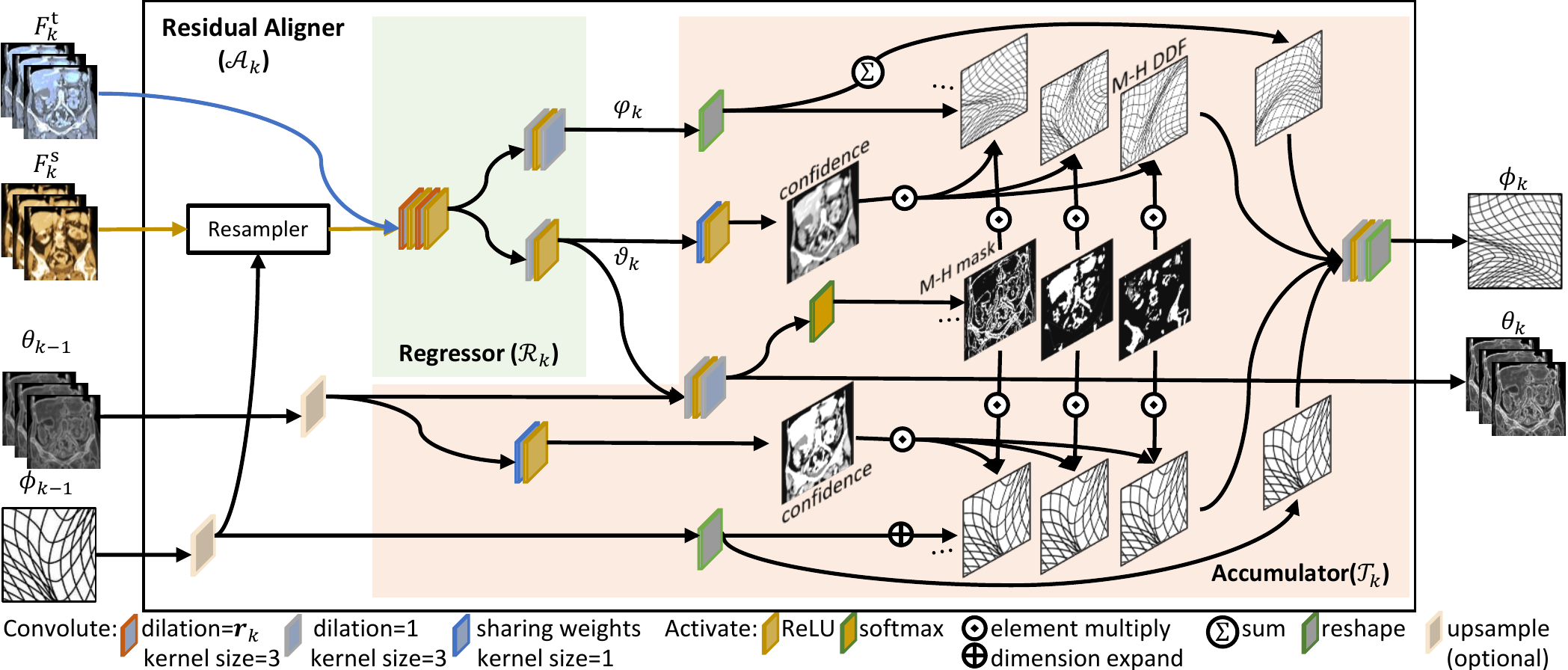}
\end{center}
\vspace{-1.5em}
\caption{The architecture of the $k^{\rm th}$ Residual Aligner (RA) module. The Regressor section regresses the residual Multi-Head (M-H) Dense Displacement Field (DDF) $\varphi_k$ and each pixel's attribute $\vartheta_k$,while the Accumulator refines the DDF $\phi_k$ via interpolation and fusion of the M-H predictions weighted by the confidence and M-H mask.}
\label{fig:res_aligner}
\end{figure*}

\section{Residual Aligners}
\label{sec:res_aligner}
The RA module as shown in Fig.~\ref{fig:res_aligner} aims to establish spatial transform $\phi$ between two images via recursively warping feature map of one towards the others, with extra attributes map $\theta$ comparing with Eq.~\eqref{equ:res_align} to restore the auxiliary information related to alignment of each pixel:
\begin{equation}
\left\{
\begin{array}{cc}
(\phi_{k},\theta_{k})=\mathcal{T}_k(\varphi_{k},\vartheta_{k},\phi_{k-1},\theta_{k-1})\\
(\varphi_{k},\vartheta_{k})=\mathcal{R}_k(\phi_{k-1}(\textbf{\textit{F}}_k^{\rm s}),\textbf{\textit{F}}_k^{\rm t})\\
\end{array}
\right.
\end{equation}
for the RA modules cascade number $k=1,...,K$. 
The $k^{\rm th}$ RA module first takes the input feature maps from source images and target images $\textbf{\textit{F}}_k^{\rm s}$, $\textbf{\textit{F}}_k^{\rm t}\in\mathbb{R}^{c_k\times n_k}$ and use the Regressor $\mathcal{R}_k$ to regress a $m$-head residual DDF $\varphi_{k}\in\mathbb{R}^{dm\times n_{k}}$ and the incremental attributes $\vartheta_{k}\in\mathbb{R}^{m\times n_{k}}$ (Sec.~\ref{sec:regress}). Then the Accumulator Network $\mathcal{T}$ computes the confidence and M-H masks, describing the prediction's reliability and the semantic properties of each pixel, and fuse the $m$-head DDF weighted based on the attribute maps $\theta_{k}\in\mathbb{R}^{m\times n_{k}}$ as in Sec.~\ref{sec:accumulator}.
The warping function performed by the resampler is implemented following the work \cite{jaderberg2015spatial}. 


\subsection{Regressor}
\label{sec:regress}
The function of Regressor $\mathcal{R}_k$ in RAN is to regress the residual transform $\varphi_{k}$ between the target feature map $\textbf{\textit{F}}_k^{\rm t}$ and the source feature map warped by the previous alignment $\phi_{k-1}(\textbf{\textit{F}}_k^{\rm s})$, with the incremental attribute map $\vartheta_k$ to restore the auxiliary information for the inter-scale refinement in the coarse-to-fine registration.
As shown in Fig.~\ref{fig:res_aligner}, Regressor concatenates the input feature maps and feeds them into the subsequent series of dilated convolution and activation layers. Referring to Sec.~\ref{sec:motion-aware}, the dilation rate vector $\textbf{\textit{r}}_k$ is set to enlarge the capture range of alignment and raise the feature resolution as introduced in Sec.~\ref{sec:net_design}.
Then two shallow convolution networks are respectively used to predict the M-H DDF and the incremental attributes raised from this level's alignment.

\subsection{Accumulator}
\label{sec:accumulator}
The task of Accumulator $\mathcal{T}_k$ is to refine the DDF with the previous coarse DDF by interpolating and fusing those spatial transform representations from varying scales and different heads in terms of the contextual information, such as the alignment reliability of the neighbouring pixels and their semantic attributes. The calculation of Accumulator shown in Fig.~\ref{fig:res_aligner} can be written as:
\begin{equation} 
\label{equ:accumulator}
\left\{
\begin{array}{cc}
    \phi_k=\mathcal{C}^{\rm 4}([{\varphi}'_{k},\sum_{\{m\}}({\varphi}_{k}),{\phi}'_{k-1},{\phi}_{k-1}])\\
    \theta_k=\mathcal{C}^{3}([\vartheta_k,\theta_{k-1}])\\
\end{array}
\right.
\end{equation}
where $\phi_{k-1}$ and $\varphi_k$ are the weighted DDF and residual DDF:
\begin{equation} 
\label{equ:weight_phi}
\left\{
\begin{array}{cc}
    {\phi}'_{k-1}=\mathcal{C}^{2}(\phi_{k-1}\otimes{{\rm softmax}(\theta_k)})\odot\mathcal{C}^{1}(\theta_{k-1})\\
    {\varphi}'_{k}=\mathcal{C}^{2}({\varphi}_k\odot{{\rm softmax}(\theta_k)})\odot\mathcal{C}^{1}(\vartheta_k)\\
\end{array},
\right.
\end{equation}
$\sum_{\{m\}}:\mathbb{R}^{d\times{m}\times{n_k}}\to\mathbb{R}^{d\times{n_k}}$ denotes the head-dimension sum, $\otimes:\mathbb{R}^{d\times{n_k}}\times\mathbb{R}^{{m}\times{n_k}}\to\mathbb{R}^{d\times{m}\times{n_k}}$ denotes the tensor product, $\odot:\mathbb{R}^{d\times{m}\times{n_k}}\times\mathbb{R}^{{1}\times{n_k}}\to\mathbb{R}^{d\times{m}\times{n_k}}$ denotes the element-wise product for the last two dimensions.
Here $\mathcal{C}^{1},\mathcal{C}^{2},\mathcal{C}^{3},\mathcal{C}^{4}$ are fitted by convolution networks with activation layers, respectively for the mapping of confidence weight projection, interpolation, attribute fusion and the DDF fusion.

\subsubsection{Confidence of Correspondence}
Simple composition of DDFs from different levels \cite{de2019deep,xu2021f3rnet} could accumulate errors at the points which failed in previous alignment. Thus, the confidence values are respectively quantified by $\mathcal{C}^{1}(\vartheta_{k}),\mathcal{C}^{1}(\theta_{k-1})$ in Eq.~\eqref{equ:weight_phi} for residual M-H DDF $\varphi_k$ and previous DDF $\phi_{k-1}$ to {weight the following filtering for interpolation with neighbouring prediction value}. 
Here the confidence is implicitly regressed from $\vartheta_{k}$ and $\theta_{k-1}$ (contrary to the confidence of occlusion probability in \cite{li2021revisiting}) with general representation aiming to provide higher accuracy. 

\subsubsection{Multi-Head Mechanism}
Inspired by the Multi-Head attention \cite{vaswani2017attention}, the corresponding M-H mask are regressed by ${{\rm softmax}(\theta_k)}$ to extract the varying motion patterns of multiple objects from the different candidate predictions at M-H residual DDF as shown in Eq.~\eqref{equ:weight_phi}.
This process could be regarded as the combination for the optimal transformation selected by the M-H masks, with preserving discontinuities in the DDF and the trend of motions \cite{heinrich2013edge}.




    



\section{Experiments}
\label{sec:experiments}

\subsection{Datasets}
We evaluated the RAN on unsupervised deformable registration on two public available datasets with segmentation annotations on 9 small organs in abdomen CT and lung in chest CT:
\\
\\
\textbf{\textit{Unpaired abdomen CT}}: The dataset is provided by~\cite{dalca2020learn2reg}. 
The ground truth segmentations of spleen, right kidney, left kidney, esophagus, liver, aorta, inferior vena cava, portal, splenic vein, pancreas of all scans are provided.
. The inter subject registration of the abdominal CT imaging is considered as challenging due to large inter-subject variations and great variability in organ volume, from 10 milliliters (esophagus) to 1.6 liters (liver). 
Each volume is resized to $2\times2\times2mm^3$ in the pre-processing. From totally 30 subjects, 23 and 7 are respectively used for training and testing, for 506 and 42 different pairing cases.
\\
\\
\textbf{\textit{Unpaired chest (lung) CT}}: The dataset is provided by~\cite{hering_alessa_2020_3835682}. The CT scans are all acquired at the same time point of the breathing cycle and here we performe inter-subject registration. 
The scanner is a Philips Brilliance 16P with a slice thickness of 1.00 mm and slice spacing of 0.70 mm. Pixel spacing in the X-Y plane varies from 0.63 to 0.77 mm with an average value of 0.70 mm. The ground truth lung segmentation of all scans are provided. Each volume is resized to $1\times1\times1mm^3$ in the pre-processing. From the total of 20 subjects, 12 and 8 are respectively used for training and testing, for 132 and 56 different pairing cases.
    
    
    
    
    
    
    


\subsection{Training details}
We normalize the input image within 0-1 range and augment the training data by randomly cropping input images during training. 
For the experiments on inter-subject registration of abdomen and lung CT, the models are first pre-trained for 50k iteration on synthetic DDF combining rigid spatial transformation and deformation (detail in Appendix).
Then the models are trained on real data for 100k iterations with the loss function:
\begin{equation}
\mathcal{L}={\mathcal{D}(\textbf{\textit{I}}^{\rm t}-\phi(\textbf{\textit{I}}^{\rm s}))}+\lambda{{\|\nabla\phi\odot{{\rm e}^{-{\|\nabla\textbf{\textit{I}}^{\rm t}\|}_2^2}}\|}_2^2}
\end{equation}
where normalized cross correlation and mean squared error are used in abdomen and lung CT respectively for $\mathcal{D}$ following \cite{balakrishnan2019voxelmorph}.
The whole training takes one week, including the data transfer, pretraining and fine-tuning. With a training batch size of 3, the initial learning rate is 0.001. The model was end-to-end trained with Adam optimizer.


\subsection{Implementation and Evaluation}
\textbf{\textit{Implementation}}:
The code for inter-subject image registration tasks were developed based on the framework of \cite{balakrishnan2018unsupervised} in Python using Tensorflow and Keras. It has been run on Nvidia Tesla P100-SXM2 GPU with 16GB memory, and Intel(R) Xeon(R) Gold 6126 CPU @ 2.60GHz. 
The backbone feature pyramid network we used is U-net \cite{ronneberger2015u} based on residual structure \cite{he2016deep} with four downsampling blocks and four upsampling blocks. Since the most motion difference ranges between 0\--15 as shown in following Fig.~\ref{fig:disp_diff}, two models RAn$_3$ and RAn$_4^+$ are thus selected as our representative models with $q=3,4$ as suggested by the effect in Fig.~\ref{fig:fcn}(d). The detail of those structures are illustrated in Appendix.
\\
\\
\textbf{\textit{Comparison}}:
We compared Residual Aligner Network with the relevant state-of-the-art methods. The Voxelmorph \cite{balakrishnan2019voxelmorph} is adopted as the representative method of direct regression (DR). The composite network combing CNN (Global-net) and U-net (Local-net) following to \cite{hu2018weakly}, as well as recursive cascaded network \cite{zhao2019recursive} are also adopted into the framework as the relevant baselines representing multi-stage (MS) networks. 
Dual-stream Pyramidal network (DPRn) \cite{hu2019dual} is selected as the baseline for feature pyramidal (FP) networks. Additionally, we also replace RA-module (in Fig.~\ref{fig:ran}) with cross attention (Attn) \cite{sun2021loftr} to compare the performance at module-level. 
\\
\\
\textbf{\textit{Evaluation metrics}}:
Following \cite{de2019deep}, we calculate the Dice Coefficient Similarity (DSC), Hausdorff Distance (HD), and Average Surface Distance (ASD) on annotated mask for the performance evaluation of nine organs in abdomen CT and one organ (lung) in chest CT, with the negative number of Jacobian determinant in tissues' region (detJ) for rationality evaluation on prediction. 
The model size, computation complexity and running time for comparison with previous methods on inter-subject registration of lung and abdomen are shown in Tab.~\ref{tab:result_abdomen_lung}. 
\begin{table}[h!]
\caption{Avg of Dice Similarity Coefficient (DSC), Hausdorff Distance (HD), Average Surface Distance (ASD) and negative number of Jacobian determinant in tissues' region (detJ) for unsupervised inter-subject registration of abdomen and chest CT using the Voxelmorph (VM1) \cite{balakrishnan2019voxelmorph} and its enhanced version with double channels (VM2), convolution networks cascaded with U-net (Cn+Un) \cite{hu2018weakly}, 5-recursive cascaded network based on the structure of VM1 and VM2 (RCn1,RCn2) as described in \cite{zhao2019recursive}, Cross Attention \cite{vaswani2017attention} w/ Feature Pyramid (CA/P), Dual-stream pyramidal registration network (DPRn) \cite{hu2019dual}, our RA network 
with $q=3,4$ (RAn$_3$,RAn$_4^+$), with different registration (reg.) types and varying Parameter Number (\#Par), Float Operations (FLOPs), and Time cost Per Image (TPI).}
\vspace{-1.5em}
\label{tab:result_abdomen_lung}
\begin{center}
\centering
\begin{tabular}{ |c|c|cccc|cccc|ccc| }
\hline
\multirow{3}{*}{model} &\multirow{2}{*}{reg. }&\multicolumn{4}{c|}{abdomen (9 organs)}&\multicolumn{4}{c|}{chest (lung)}&\multicolumn{3}{c|}{efficiency}\\
&\multirow{2}{*}{type}
&\cellcolor[RGB]{255,170,170}DSC$\uparrow$ &\cellcolor[RGB]{153,204,255}HD$\downarrow$  &\cellcolor[RGB]{153,204,255}ASD$\downarrow$
&\cellcolor[RGB]{153,204,255} det{J}$\downarrow$
& \cellcolor[RGB]{255,170,170}DSC$\uparrow$ &\cellcolor[RGB]{153,204,255}HD$\downarrow$  &\cellcolor[RGB]{153,204,255}ASD$\downarrow$
&\cellcolor[RGB]{153,204,255} det{J}$\downarrow$
&\cellcolor[RGB]{153,204,255}\#Par$\downarrow$&\cellcolor[RGB]{153,204,255}FLOPs$\downarrow$&\cellcolor[RGB]{153,204,255}TPI$\downarrow$
\\
&&\cellcolor[RGB]{255,170,170}(\%)&\cellcolor[RGB]{153,204,255}(mm)&\cellcolor[RGB]{153,204,255}(mm)&\cellcolor[RGB]{153,204,255}(e{3})&\cellcolor[RGB]{255,170,170}(\%)&\cellcolor[RGB]{153,204,255}(mm)&\cellcolor[RGB]{153,204,255}(mm)&\cellcolor[RGB]{153,204,255}(e{3})
&\cellcolor[RGB]{153,204,255}(e6)&\cellcolor[RGB]{153,204,255}(e9)&\cellcolor[RGB]{153,204,255}(sec)
\\

\hline
Initial&-
&30.9&49.5&16.04&-    
&61.9&41.6&15.86&-
& -&- &-
\\
VM1 &DR
&44.7&43.8&9.24&2.23
&84.0&32.9&6.38&5.94
& 0.36& 34.2& 0.23
\\
VM2 &DR
&51.9&45.0&8.40&4.03
&88.8&32.0&5.02&15.58
& 1.42 & 69.6& 0.25
\\
CA/P&Attn
&47.6&43.8&8.77&3.85
& 84.7&28.9& 5.75&2.67
& 0.58 & 114.5 &0.41
\\
Cn+Un &MS
&53.6&44.6&7.84&4.13
&91.1&29.7& 3.84&4.23
&2.11&94.7&0.36
\\ 
RCn1 &MS
&55.6&44.9&7.79&2.91
&89.8&33.1&4.68&5.68
&0.36  & 219.2&0.44
\\  
RCn2 &MS
&59.5&44.1&6.95&\textbf{1.36}
&\textbf{93.7}&29.1&3.04&\textbf{1.66}
& 1.42& 308.7& 0.45
\\  
DPRn&FP
&53.9&57.1&8.18&4.28
&88.4&29.9&4.48&3.46
& 0.62 &  82.1 &0.46
\\
\rowcolor[RGB]{230,230,230} {RAn$_{3}$} &MA
&54.2&43.8&7.74&3.48
&{{93.5}}&\textbf{26.3}&\textbf{3.01}&4.05
&0.72  &132.1  & 0.48
\\
\rowcolor[RGB]{230,230,230} {RAn$^+_{4}$} &MA
&\textbf{61.7}&\textbf{40.8}&\textbf{6.51}&1.55
&{91.6}&29.2&{3.84}&3.17
&0.75  &229.7  &  0.56
\\
 \hline
\end{tabular}
\end{center}
\end{table}
\begin{table}[h!]
\caption{Ablation study on RA module by inter-subject image registration of abdomen CT and lung CT using, with varying setting of head number $m$, motion-aware pattern of feature maps ($q=0,3,4$) and confidence weights (CW).}
\vspace{-1.5em}
\label{tab:result_ablation}
\begin{center}
\centering
\begin{tabular}{ |c|ccc|cccc|cccc|ccc| }
\hline
\multirow{3}{*}{model} &\multicolumn{3}{c|}{setting}&\multicolumn{4}{c|}{abdomen (9 organs)}&\multicolumn{4}{c|}{chest (lung)}&\multicolumn{3}{c|}{efficiency}\\
&\multirow{2}{*}{CW}&\multirow{2}{*}{MH}&\multirow{2}{*}{$q$}
& \cellcolor[RGB]{255,170,170}DSC$\uparrow$ &\cellcolor[RGB]{153,204,255}HD$\downarrow$  &\cellcolor[RGB]{153,204,255}ASD$\downarrow$&\cellcolor[RGB]{153,204,255} det{J}$\downarrow$& \cellcolor[RGB]{255,170,170}DSC$\uparrow$ &\cellcolor[RGB]{153,204,255}HD$\downarrow$  &\cellcolor[RGB]{153,204,255}ASD$\downarrow$&\cellcolor[RGB]{153,204,255} det{J}$\downarrow$
&\cellcolor[RGB]{153,204,255}\#Par$\downarrow$&\cellcolor[RGB]{153,204,255}FLOPs$\downarrow$&\cellcolor[RGB]{153,204,255}TPI$\downarrow$
\\
&&&&\cellcolor[RGB]{255,170,170}(\%)&\cellcolor[RGB]{153,204,255}(mm)&\cellcolor[RGB]{153,204,255}(mm)&\cellcolor[RGB]{153,204,255}(e3)&\cellcolor[RGB]{255,170,170}(\%)&\cellcolor[RGB]{153,204,255}(mm)&\cellcolor[RGB]{153,204,255}(mm)&\cellcolor[RGB]{153,204,255}(e3)
&\cellcolor[RGB]{153,204,255}(e6)&\cellcolor[RGB]{153,204,255}(e9)&\cellcolor[RGB]{153,204,255}(sec)\\
 \hline
DPRn&&&0
&53.4&57.1&8.18&4.28
&88.4&29.9&4.48&3.46
& 0.62 &  82.1 &0.46
\\
RAn &\checkmark&&0
&53.9&46.0&8.03&2.65
& 90.7&30.3&3.74&9.61
& 0.68 & 83.9&0.41
\\
RAn&\checkmark&&4
&56.4&{44.8}&7.48&2.66
&{92.1}&28.1&{3.42}&6.87
& 0.71 & 170.5& 0.56
\\
{RAn$_{0}$} &\checkmark&\checkmark&0
&53.3&44.0&7.98&2.64
& 92.5&{28.9}&3.34&3.74
& 0.71 & 101.3&0.47 
\\
{RAn$_{3}$} &\checkmark&\checkmark&3
&54.2&43.8&7.74&3.48
&{\textbf{93.5}}&\textbf{26.3}&\textbf{3.01}&4.05
&0.72  &132.1  & 0.48
\\
{RAn$_{4}$} &\checkmark&\checkmark&4
&56.1&44.2&7.66&2.46
&{91.5}&26.9&{3.55}&4.06
&0.71  &192.3  &  0.54
\\
{RAn$^+_{4}$} &\checkmark&\checkmark&4
&\textbf{61.7}&\textbf{40.8}&\textbf{6.51}&\textbf{1.55}
&{91.6}&29.2&{3.84}&\textbf{3.17}
&0.75  &229.7  &  0.56
\\
 \hline
\end{tabular}
\end{center}
\end{table}
\begin{figure}[t!]
\begin{center}
\includegraphics[width=.95\linewidth]{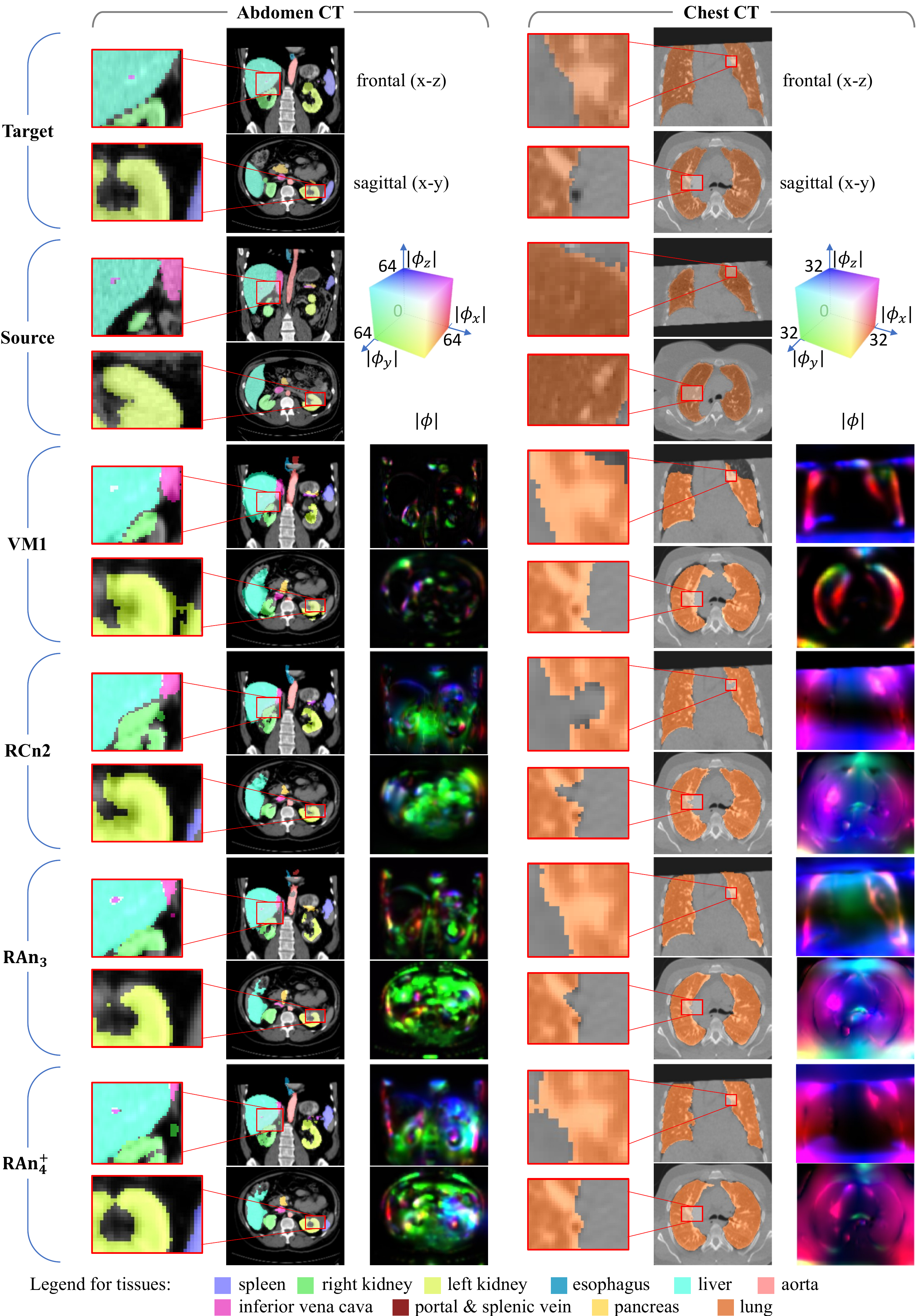}
\end{center}
\vspace{-1.5em}
\caption{Qualitative example in abdomen and chest CT shows our network achieves plausible registration, with the improvement at the areas between different organs, such as liver, inferior vena cava and right kidney, as well as the edge area of the lung.}
\label{fig:qual_result}
\end{figure}

\begin{figure}[t!]
\begin{center}
\includegraphics[width=\linewidth]{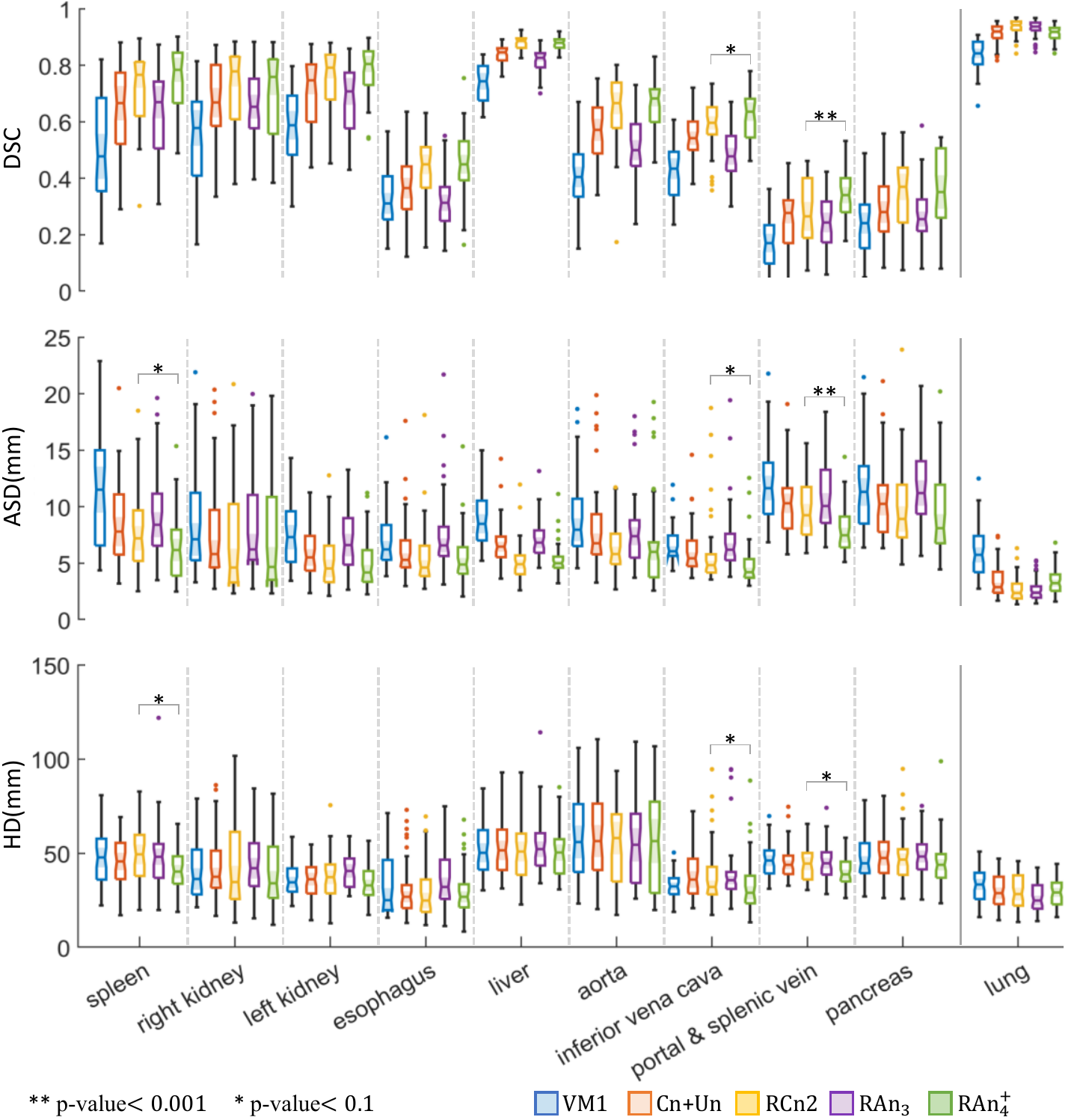}
\end{center}
\vspace{-1.5em}
\caption{RANs achieve the best registration in the veins in abdominal CT scans. The box plots of DSC, ASD and HD illustrate our networks achieve the best registration in infer vena cava and portal \& splenic vein, (sample number 42\&56 for abdomen\&chest). RANs  equipped with higher $q$ performs better on the smaller organs (c.f. RAn$_3$ with RAn$_4^+$).}
\label{fig:quan_result}
\end{figure}

\begin{figure}[th!]
\begin{center}
\includegraphics[width=\linewidth]{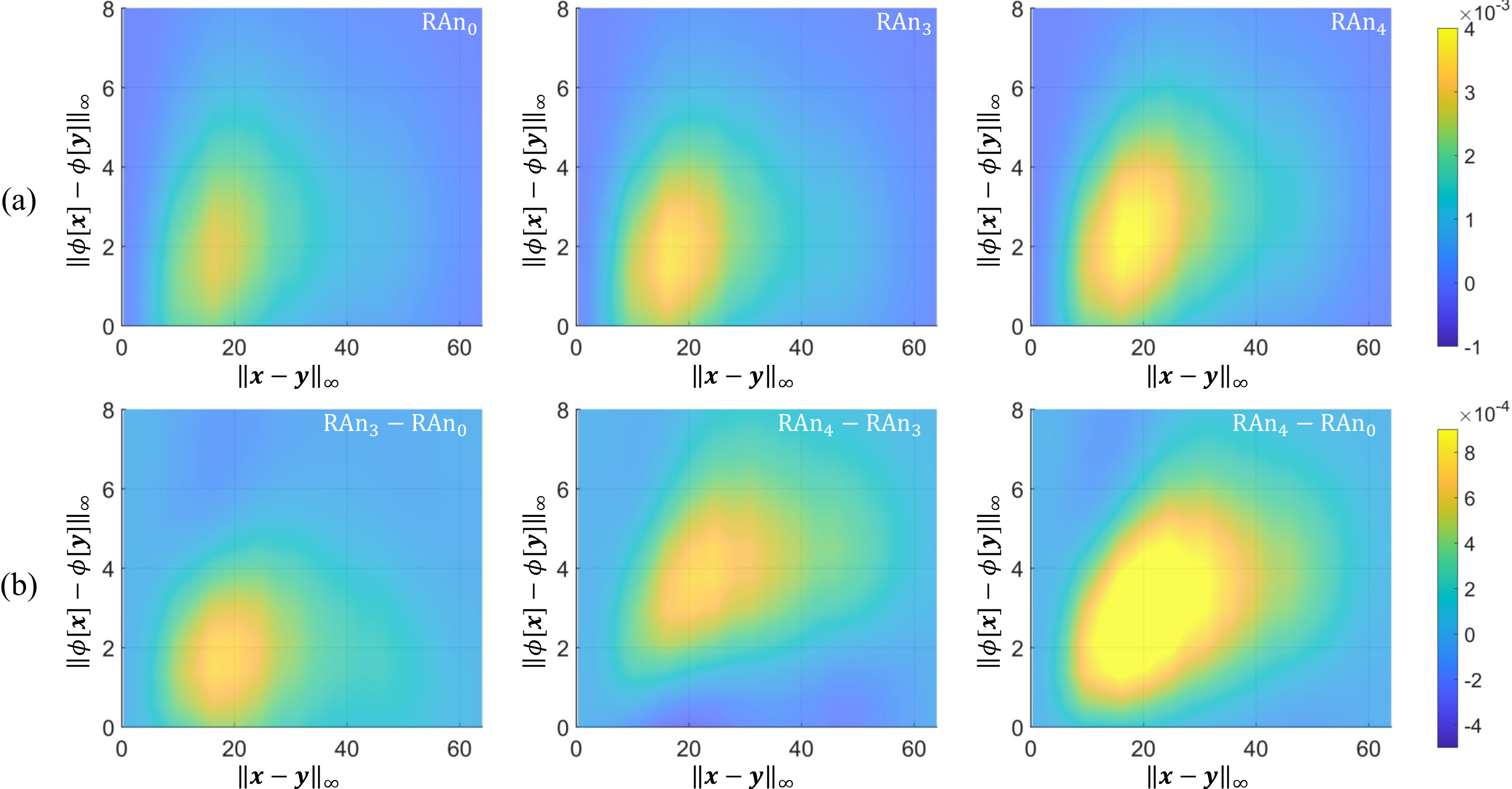}
\end{center}
\vspace{-1.5em}
\caption{(a) Probability Density Function (PDF) of the pairs of correct predicted motions by RAn$_0$, RAn$_3$ and RAn$_4$, smoothed by Gaussian filter ($\sigma$=1pix), with respect to varying Chebyshev distance (${\|\textbf{\textit{x}}-\textbf{\textit{y}}\|}_{\infty},\forall{\textbf{\textit{x}},\textbf{\textit{y}}\in\{\textbf{\textit{x}}|\textbf{\textit{L}}^{\rm t}[\textbf{\textit{x}}]=\phi(\textbf{\textit{L}}^{\rm s})[\textbf{\textit{x}}]\}}$) and varying Chebyshev distance between their motions (${\|\phi[\textbf{\textit{x}}]-\phi[\textbf{\textit{y}}]\|}_{\infty}$), and (b) the difference between each pair of PDF, validating that higher number of MA pattern layers enable network to achieve better motion separability with similar model scale, where $\textbf{\textit{L}}^{\rm s\&t}\in\{{\rm spleen},\cdots,{\rm pancreas}\}^{n}$ denote the labels on source\&target images of abdomen CT.}
\label{fig:disp_diff}
\end{figure}

\subsection{Results}
\textbf{\textit{Performance on registration}}:
The comparison between RAN with other methods on abdomen and chest CT using all 10 organs is shown in Tab.~\ref{tab:result_abdomen_lung}, and the results illustrate our network achieved one of the best performance in this task with fewer parameters and lower computational cost. The Fig.~\ref{fig:qual_result} illustrate the improvement at the area containing multi-organs and at the edges of organs.
In terms of registration types, DR networks (VM2) require more parameters for better results, and MS networks (RCn1, RCn2) need much more computation, while the FP based network (DPRn) balances between them, and our MA based RAn further improve it. 
The separate evaluation of the 9+1 organs (abdomen+chest) for five models, as shown in Fig.~\ref{fig:quan_result}, illustrates our RAn achieves best accuracy in the registration of small organs (veins) and one of the best accuracy in other organs' registration.  
\\
\\
\textbf{\textit{Ablation Study}}:
To validate the effect of each component on the performance, we also tried several combination on the confidence weight (CW), multi-heads (MH) and motion-aware pattern number ($q$) on experiments of abdomen and lung CT as shown in Tab.~\ref{tab:result_ablation}. For a fair comparison, the channel numbers are tuned to keep the trainable parameter numbers similar to each others, except RAn$_4^+$ with larger model size for higher accuracy. 
Fig.~\ref{fig:qual_result} and \ref{fig:quan_result} show our RAn$_4^+$ with $q=4$ is better than RAn$_3$ on smaller tissues' prediction but worse on larger one's (lung).
\\
\\
\textbf{\textit{Separability of the predicted motions}}:
\label{sec:pred_motion_sep}
Beside implicitly validated by the better results of higher $q$, more visual validation of MA design is illustrated in Fig.~\ref{fig:disp_diff}(a) including the probability density distributions of the pairs of correct motion prediction with varying voxel distance and motion difference for varying $q$. Based on the difference between them in Fig.~\ref{fig:disp_diff}(b), It shows RAn with higher $q$ obtain more correction hits at the left-top area, and thus the better motion separability, matching the expectation in Fig.~\ref{fig:fcn}(d) and validating the improvement by the design principle described in Sec~\ref{sec:motion-aware}.

\section{Discussion and Conclusion}
The novel RAN design is proposed based on the MA mechanism with a new RA module. It achieves the best registration of the veins in abdominal CT scans and comparable registrations with other state of the art networks in the other tissues in abdominal and chest CT with fewer parameters and less computation. 
Additionally, RANs based on MA structure achieved the improving separability of predicted motion as shown in Fig.~\ref{fig:disp_diff}, which also validate the proposed design principle on MA mechanism. 
These results demonstrate the efficiency and the potential of RAn performing on relevant tasks including multi-object registration, which could also be further applicable to other relevant computer vision tasks, such as optical flow, stereo matching and motion tracking. 



%
%
\bibliographystyle{splncs04}
\bibliography{egbib}
\end{document}